\documentstyle[a4wide,epsf,11pt,titlepage]{article}

\newcounter{nref}
\newcommand{\bbib}{%
  \renewcommand{\refname}{\large\bf References}%
  \begin{thebibliography}{99}%
  \setcounter{enumiv}{\thenref}}
\newcommand{\ebib}{%
  \end{thebibliography}%
  \setcounter{nref}{\arabic{enumiv}}}
\newcommand{\head}[3]{%
  \setcounter{nref}{0}%
  \thispagestyle{empty}%
  \section*{\LARGE\bf #1}%
  \stepcounter{section}%
  \addcontentsline{toc}{section}{#1}%
  \large\itshape%
  #2\\\vspace{0.1pt}\\%
  #3%
  \normalsize\upshape%
  \bigskip}

\begin{document}


\head{Turbulent Combustion in Type Ia Supernova Models}
     {F.K.\ R\"opke, W.\ Hillebrandt}
     {Max-Planck-Institut f\"ur Astrophysik,
     Karl-Schwarzschild-Str.\ 1, D-85741 Garching, Germany }

\subsection*{Abstract}

We review the astrophysical modeling of type Ia supernova explosions
and describe numerical methods to implement numerical simulations of
these events. Some results of such simulations are discussed.

\subsection{Astrophysical and numerical models}

Type Ia supernovae (SNe Ia) are among the brightest and most energetic
explosions observed in the Universe. For a short time they can
outshine an entire galaxy consisting of some hundred billions of
stars. Assuming that
SNe Ia originate from a single stellar object, only two sources of
explosion energy come into consideration: the gravitational binding
energy of the star and its nuclear energy. Since for the particular
class of SNe Ia no compact object is found in the remnant, they are
usually associated with thermonuclear explosions of white dwarf (WD)
stars consisting of carbon and oxygen. The currently favored
astrophysical model assumes it to be part
of a binary constellation and to accrete matter from the companion
until it comes close to the limiting Chandrasekhar mass. At this
stage, the central density of the WD reaches values at which nuclear
burning of carbon towards heavier elements ignites. After a
simmering phase of several hundreds of years, a thermonuclear runaway
in a tiny region close to the center leads to the formation
of a thermonuclear flame. 

The astrophysical interest in SNe Ia is -- among other things --
founded on their relevance for cosmology. On the basis of an empirical
calibration relating their peak luminosities with the shapes of their
lightcurves they are a suitable tool to determine cosmological
distances. The geometrical survey of the Universe performed in this way
led to one of the greatest surprises of modern astrophysics pointing to
the fact that the Universe is predominantly made of a so far unknown
``dark energy'' component. SNe Ia distance measurements may in the
future possibly contribute to the determination of the equation of
state of this dark energy. However, the empirical calibration applied
here urgently calls for a theoretical explanation and ongoing SN Ia
cosmology projects crucially depend on increasing the accuracy of the
measurements by getting a handle on the systematic errors. This is
achievable only on the basis of a better understanding of the
mechanism of SN Ia explosions.

To this end, we attempt to model SN Ia explosions from ``first
principles'' in conjunction with detailed comparison with observations
of nearby objects. The goal is to construct numerical models as
parameter-free as possible.

Such a SN Ia explosion model has to describe the propagation of the
thermonuclear flame from the WD's center outwards. Hydrodynamics in
principle allows for two distinct modes here. One is the so-called
deflagration mode, in which the subsonic flame is mediated by the
thermal conduction of the degenerate electrons, and the other is a
supersonic detonation in which the flame is driven by sound waves.

A prompt detonation has been ruled out as a valid model for SNe Ia,
since the entire star is incinerated with sound speed here. Therefore the
material has no time to pre-expand and is burned at high densities
where the nuclear reactions terminate in iron group elements. This is
in disagreement with observations showing that intermediate mass
elements need to be produced as well. Hence, the flame must start out
subsonically in the deflagration mode. However, a
laminar deflagration flame is much too slow to release sufficient
energy to explode the star. The main issue of SN Ia models is thus to
identify mechanisms to accelerate the flame propagation.

This is the point where turbulence comes into play. The interaction of
the flame with turbulent motions defines burning in SNe Ia as a
problem of turbulent combustion. The flame propagating from the center
of the star outwards produces an inverse density stratification in the
gravitational field of the WD leaving light and hot nuclear ashes
behind while the fuel in front of it is dense and cold. The resulting
Rayleigh-Taylor instability leads to the formation of burning bubbles
that buoyantly rise into the fuel. The shear flows at the interfaces
of these bubbles are characterized by a Reynolds number of about
$10^{14}$ and the Kelvin-Helmholtz instability generates turbulent
eddies. These decay in a turbulent energy cascade and the flame
interacts with eddies on a wide range of scales. In this way, the flame
becomes corrugated and its surface area is enlarged. This enhances the
net burning rate and accelerates the flame propagation. A later
transition of the flame propagation mode is still hypothetical and not
further discussed here. 

For a numerical implementation of the deflagration SN Ia model, the
scale down to which the flame interacts with turbulent motions has to
be considered. This is the so-called Gibson scale, at which turbulent
velocity fluctuations of the cascade reach values comparable with the
laminar flame speed. At the beginning of the explosion (the WD star
has a radius of about $2000 \, \mathrm{km}$ and ignites inside the
first $\sim$$100 \, \mathrm{km}$), the Gibson scale is of the order of
$10^4 \, \mathrm{cm}$. The flame width, however, is less than a
millimeter. Due to this huge scale separation, turbulent eddies
interact with the flame only in a kinematic way but leave the internal
flame structure unaffected. Thus, burning proceeds in the so-called
flamelet regime of turbulent combustion for most parts of the
explosion process. With three-dimensional simulations on scales of the
WD star, it is possible to reach resolutions down to less than a kilometer. Of
course, these simulations need to take into account effects of
turbulence on smaller (unresolved) scales, which is implemented via a
sub-grid
scale model (cf.\ the contribution of W.~Schmidt et
al.). Complementary small-scale simulations are provided to test the
assumptions of flame propagation around and below the Gibson scale. 

One has to keep in mind, however, that the explosion process takes
place on an expanding background. Due to the energy release, the WD
expands. With lower fuel densities, the flame structure broadens and
the laminar flame speed decreases
\cite{roepke.1}. Therefore the Gibson scale becomes smaller and
eventually, in the very late phases of the 
explosion, turbulent eddies may be capable of penetrating the
flame structure so that the distibuted burning regime is entered.

The numerical implementation of the outlined SN Ia model on scales of
the WD star follows
\cite{roepke.2} in a large eddy simulation (LES)
approach. The resolved hydrodynamics is described by the
\textsc{Prometheus} implementation \cite{roepke.3} of a higher-order
Godunov scheme. Turbulence on unresolved scales is taken into account
with a sub-grid scale model. Seen from scales of the WD, the flame
appears as a sharp discontinuity separating the fuel from the
ashes. Its propagation is modeled via the level set method
\cite{roepke.4}, where the flame velocity is set by the
physics of the flamelet regime. Here, flame propagation completely
decouples from the microphysics of the burning and is determined by
the turbulent velocity fluctuations on the grid scale which are known
from the sub-grid scale model. The nuclear reactions are implemented
in the simplified approach of \cite{roepke.5}.

\subsection{Results}

\begin{figure}[t]
  \centerline{\epsfxsize=0.717 \textwidth\epsffile{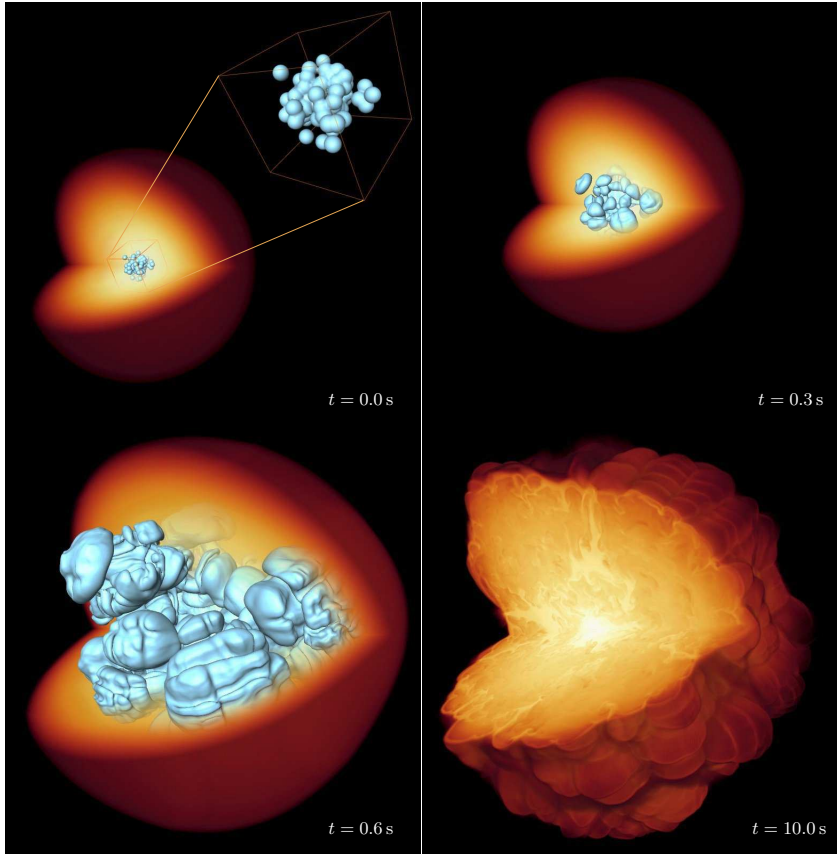}}
  \caption{Snapshots from a SN Ia explosion simulation. The WD star
    is indicated by the volume rendering of the logarithm of the
    density and the isosurface corresponds to the flame front. The
    snapshot at $t = 10.0 \, \mathrm{s}$ is not on scale with the
    other images. (Simulation from \cite{roepke.6})}
  \label{roepke.fig1}
\end{figure}

Numerical simulations on the basis of the outlined model have been
shown to lead to explosions of the WD star. A flame ignited near the
center of the star (cf.\ top left panel of Fig.~\ref{roepke.fig1})
develops the typical ``mushroom''-like features due
to buoyancy instabilities (cf.\ top right panel of
Fig.~\ref{roepke.fig1}). It becomes increasingly wrinkled and the
generated turbulence accelerates the flame propagation. In this way,
the flame incinerates considerable fractions of the material 
(cf.\ bottom left panel of Fig.~\ref{roepke.fig1}) and the
energy release is sufficient to gravitationally unbind the WD
star. A snapshot of the density structure of the remnant after the
burning has ceased is shown in the bottom right panel of
Fig.~\ref{roepke.fig1}, where the imprints of turbulent burning are
clearly visible. The most vigorously exploding model so far
released about $7 
\times 10^{51}\, \mathrm{erg}$ of energy \cite{roepke.7}. Another
important global quantity to asses the explosion process is the mass
of produced $^{56}$Ni, because its radioactive dacay powers the
visible event. In the mentioned simulation, $0.4\, M_\odot$ of
$^{56}$Ni were
obtained. Both values are within the range of expectation from
observations, albeit on the low side. First synthetic light curves
have been derived from explosion simulations \cite{roepke.8} and
compare well with observations.

However, current deflagration models of SNe Ia seem to have
difficulties
reproducing observed spectra. A spectrum of the late (``nebular'')
phase at day 350 after explosion was recently derived \cite{roepke.9}
from a very simple simulation. Although reproducing the
broad iron lines of observed spectra well, it showed strong indication of
unburnt material at low velocities which is not seen in the
observations. Both features, however, share a common origin. The
rising bubbles filled with ashes distribute iron group elements over a
wide range in velocity space and thus give rise to broad iron
lines. At the same time, downdrafts in between these bubbles transport
unburnt material towards the center producing the strong oxygen and
carbon lines which are absent in the observations.

This problem may in part be attributed to the simplicity of the
underlying explosion simulation. It was performed on only one octant
of the star with rather low resolution and the flame was ignited in a
very artificial
shape. Nonetheless, it seems likely that physical ingredients are still
missing in the explosion model. In particular, burning at late phases
was ignored as yet. Fuel consumption was ceased when the flame
reached densities of unburnt material below $10^7 \, \mathrm{g} \,
\mathrm{cm}^{-3}$, because the distributed burning regime is expected
to be entered here. A recent
approach \cite{roepke.10}, however, modeled the transition between the
turbulent burning regimes by assuming 
flamelet scaling for the flame propagation velocity above this density
threshold and by applying Damk\"ohler's limit for the thin reaction
zone regime \cite{roepke.11} below. The result strongly supports the
conjecture that an implementation of burning at low densities may help
to cure current problems of the deflagration SN Ia model.



\bbib
\bibitem{roepke.1} F.X.~Timmes and S.E.~Woosley, ApJ
    {\bf 396} (1992) 649.
\bibitem{roepke.2} M.~Reinecke, W.~Hillebrandt, J.C.~Niemeyer,
   R.~Klein, and A.~Gr{\"o}bl, A\&A
    {\bf 347} (1999) 724.
\bibitem{roepke.3} B.A.~Fryxell and E.~M{\"u}ller, MPA Green Report
    {\bf 449} (Max-Planck-Institut f{\"ur} Astrophysik, Garching, 1989).
\bibitem{roepke.4} S.~Osher and J.A.~Sethian, J. Comp. Phys.
    {\bf 79} (1988) 12.
\bibitem{roepke.5} M.~Reinecke, W.~Hillebrandt, and J.C.~Niemeyer, A\&A
    {\bf 386} (2002) 936.
\bibitem{roepke.6} F.K.~R{\"o}pke and W.~Hillebrandt, A\&A
    {\bf 431} (2005) 635.
\bibitem{roepke.7} C. Travaglio, W.~Hillebrandt, M.~Reinecke, and F.-K.~Thielemann, A\&A
    {\bf 425} (2004) 1029.
\bibitem{roepke.8} E.~Sorokina and S.~Blinnikov, in ``From Twilight to
  Highlight: The Physics of Supernovae'' 
    (Springer-Verlag, Berlin Heidelberg, 2003) 268.
\bibitem{roepke.9} C.~Kozma, C.~Fransson, W.~Hillebrandt,
  C.~Travaglio, J.~Sollerman, M.~Reinecke, F.K.~R{\"o}pke, and
  J.~Spyromillio, submitted to A\&A (astro-ph/0504299).
\bibitem{roepke.10} F.K.~R{\"o}pke and W.~Hillebrandt, A\&A
    {\bf 429} (2005) L29.
\bibitem{roepke.11} G. Damk{\"o}hler, Z.\ f.\ Electrochem.\
    {\bf 46} (1940) 601.
\ebib
\end{document}